\documentstyle[epsfig]{aipproc}

\begin{document}

\newcommand{\snr}{{G266.2$-$1.2}}

\title{Nonthermal X-Ray Emission from G266.2$-$1.2 (RX~J0852.0$-$4622)}

\author{P. Slane$^1$, J. P. Hughes$^2$, R. J. Edgar$^1$, 
P. P. Plucinsky$^1$,\\
 E. Miyata$^3$, and B. Aschenbach$^4$}
\address{$^1$Harvard-Smithsonian Center for Astrophysics, Cambridge, MA 
02138\\
$^2$Rutgers University, Piscataway, NJ 08854-8019\\
$^3$Osaka University, Osaka 560-0043 JAPAN\\
$^4$Max-Planck-Institut f\"{u}r extraterrestrische Physik, Garching, Germany
}

\maketitle

\begin{abstract}
The newly discovered supernova remnant G266.2$-$1.2 (RX~J0852.0--4622), 
along the line of sight to the Vela SNR, was observed with ASCA for 120 ks.
We find that the X-ray spectrum is featureless, and well described by
a power law, extending to three the class of shell-type SNRs dominated by 
nonthermal X-ray emission. Although the presence of the Vela SNR
compromises our ability to accurately determine the column density,
the GIS data appear to indicate absorption considerably in excess of
that for Vela itself, indicating that G266.2-1.2 may be several times
more distant. An unresolved central source may be an associated neutron 
star, though difficulties with this interpretation persist.
\end{abstract}

\section*{Introduction}

\begin{figure}[t!] 
\centerline{
\epsfig{file=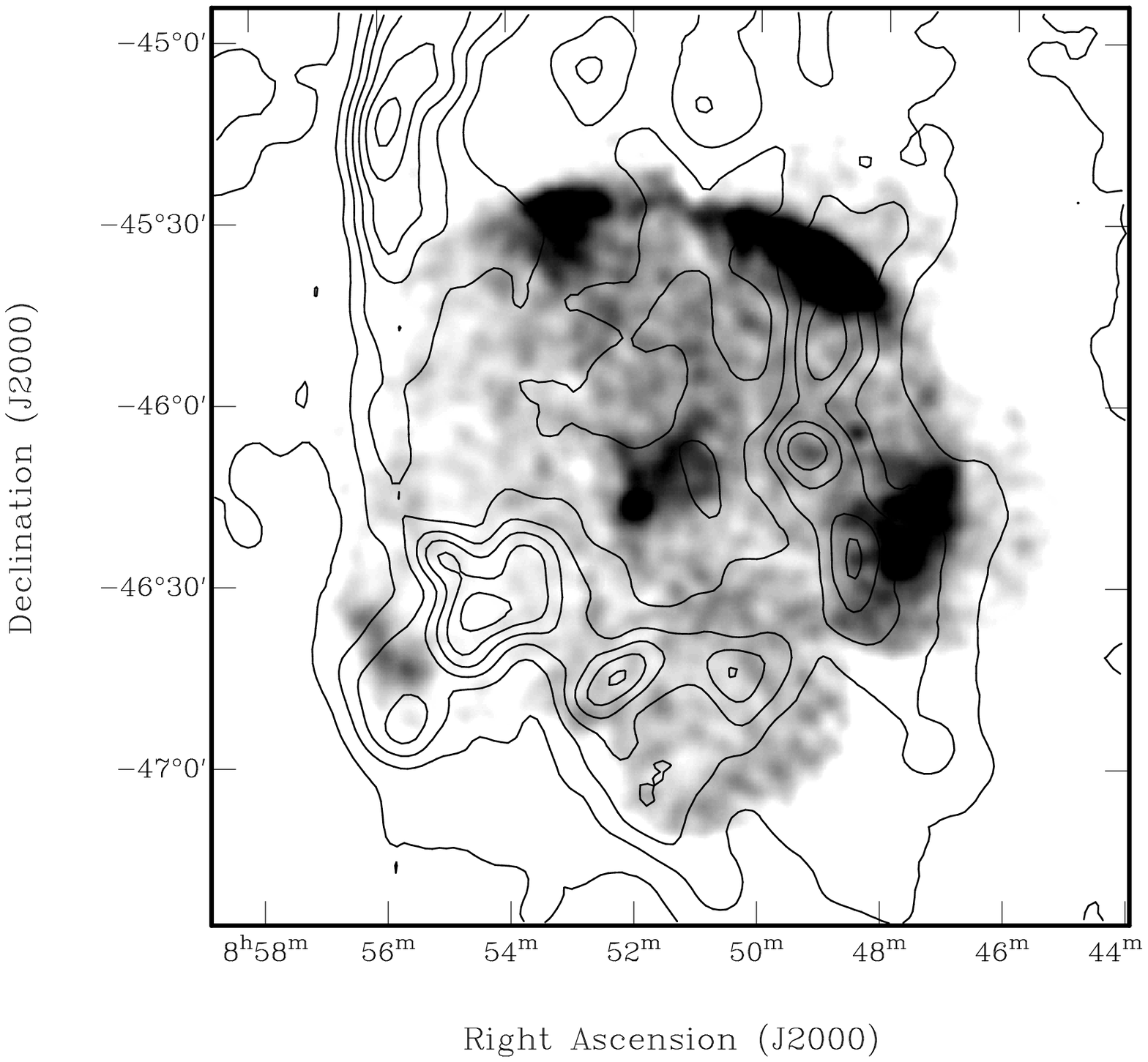,width=2.6in}\hspace{0.4in}
\epsfig{file=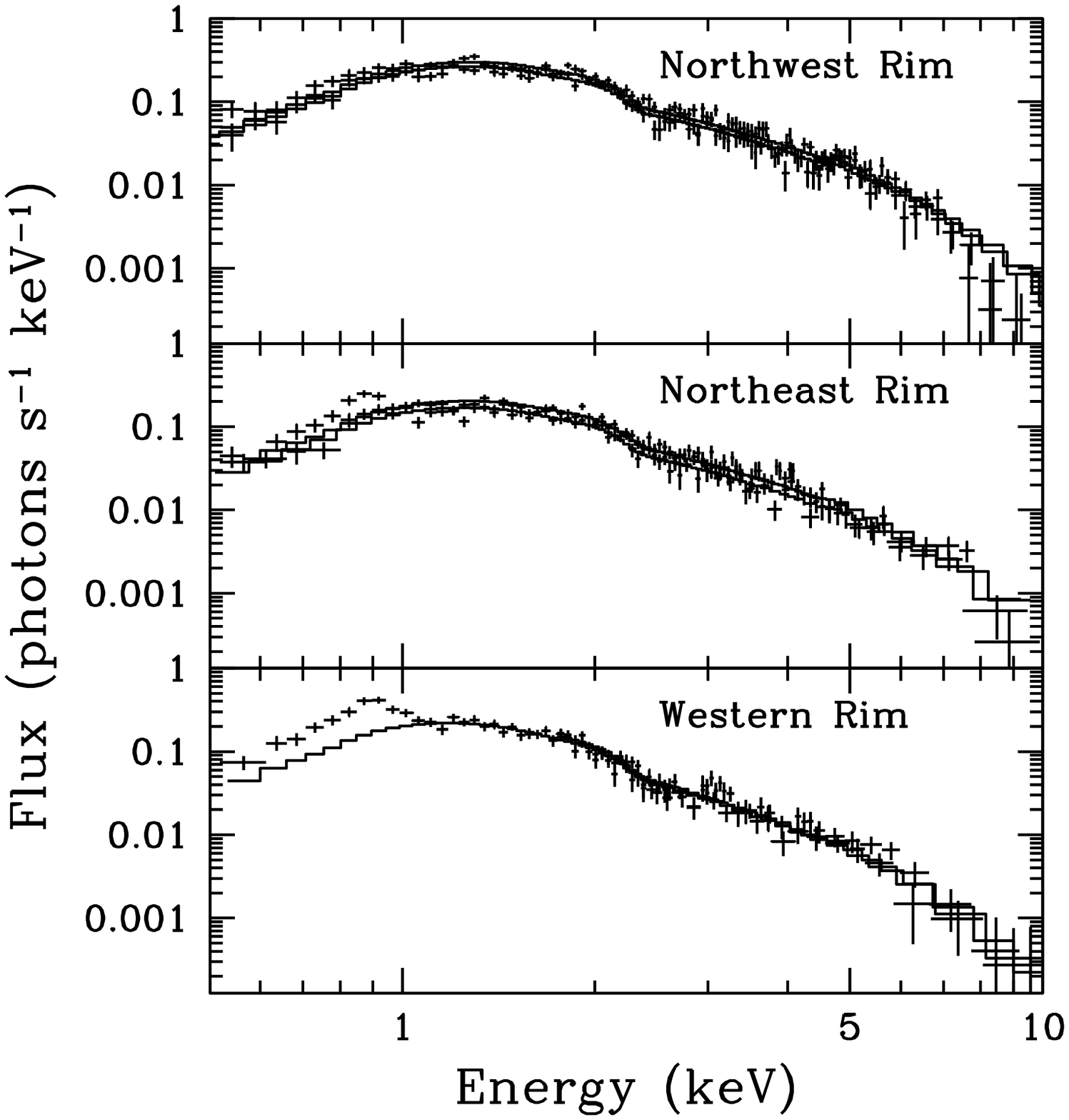,width=2.6in}}
\caption{Left: ASCA GIS image of G266.2--1.2 ($E = 0.7 - 10$~keV). The image
consists of a mosaic of 7 individual fields. Contours represent the outline
of the Vela SNR as seen in ROSAT survey data with the PSPC. Right: 
ASCA spectra from both GIS detectors for regions of G266.2--1.2.
The featureless spectrum
is well described by a power law. Excess flux at low energies is
presumably associated with soft thermal emission from the Vela SNR.} 
\label{fig1}
\vspace{-0.13in}
\end{figure}

\snr\ was discovered by Aschenbach \cite{asch1} using
data from the ROSAT All-Sky Survey. Situated along the line of sight
to the Vela SNR, the emission stands out above the soft
thermal emission from Vela only at energies above $\sim 1$~keV. 
Iyudin et al. \cite{iyud} reported the detection of $^{44}$Ti from
the source GRO~J0852-4642, which was tentatively associated with the SNR.
If correct, 
the very short $^{44}$Ti
lifetime ($\tau \approx 90$~y) would imply a very young SNR, and the
large angular size would require that the remnant be very nearby as
well. Estimates based on the X-ray diameter and $\gamma$-ray flux
of $^{44}$Ti indicate an age of $\sim 680$~y and a distance of
$\sim 200$~pc \cite{aschetal}.
The hard X-ray spectrum, if interpreted as high temperature emission
from a fast shock, would seem to support this scenario.
However, as we summarize here (see \cite{slane01}
for a detailed discussion), the X-ray emission is not from hot,
shock-heated gas; it is nonthermal.
Further, reanalysis of the COMPTEL data finds that the $^{44}$Ti detection 
is only significant at the
$2 - 4 \sigma$ confidence level \cite{schon}.
In the absence of such emission, and given that the X-ray emission
is nonthermal, the nearby distance and young age may need to be reexamined.

\section*{Observations and Analysis}
We have carried out ASCA observations of \snr\ using
7 pointings, each of $\sim 17$~ks duration. The
resulting image from the ASCA GIS is illustrated in
Fig. 1 \cite{tsun}, where we also
present GIS spectra from three regions
along the rim of the remnant: 1) the bright
NW rim; 2) the NE rim; and 3) the W rim.
Unlike the line-dominated thermal emission typical of
a young SNR, the spectra are featureless
and well described by a power law of index $\sim 2.6$. \snr\ thus joins
SN~1006 \cite{koy95} and G347.3$-$0.5 \cite{koy97,slane97} as a shell-type
SNR dominated by nonthermal emission in X-rays. Weaker nonthermal
components are observed for other SNRs as well, indicating the 
shock-acceleration of particles to energies of order
$10-100$~TeV.

The best-fit spectral parameters for \snr\ yield 
$N_H \sim (1 - 4) \times 10^{21}{\rm\ cm}^{-2}$. 
We include a soft thermal component with the column density fixed
at a low value appropriate for Vela \cite{bocc}.
The column density for the power law component 
is significantly higher than that for Vela. While simple scaling of
the column density to estimate the distance to \snr\ is clearly
rather uncertain, it would appear that the remnant is at least
several times more distant than Vela; 
improved column density measurements are of considerable importance.

\begin{figure}[t!] 
\centerline{
\epsfig{file=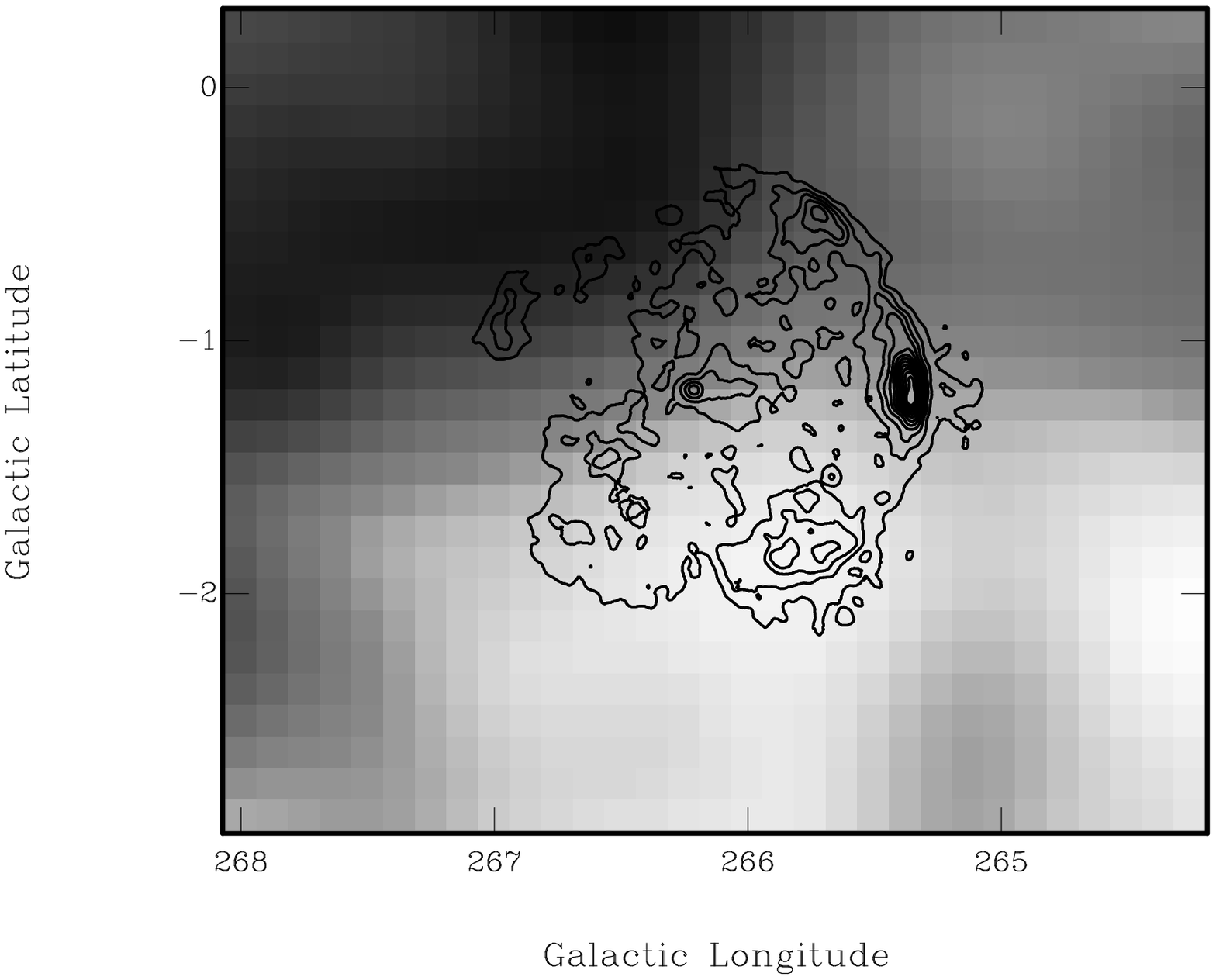,width=2.6in} \hspace{0.4in}
\epsfig{file=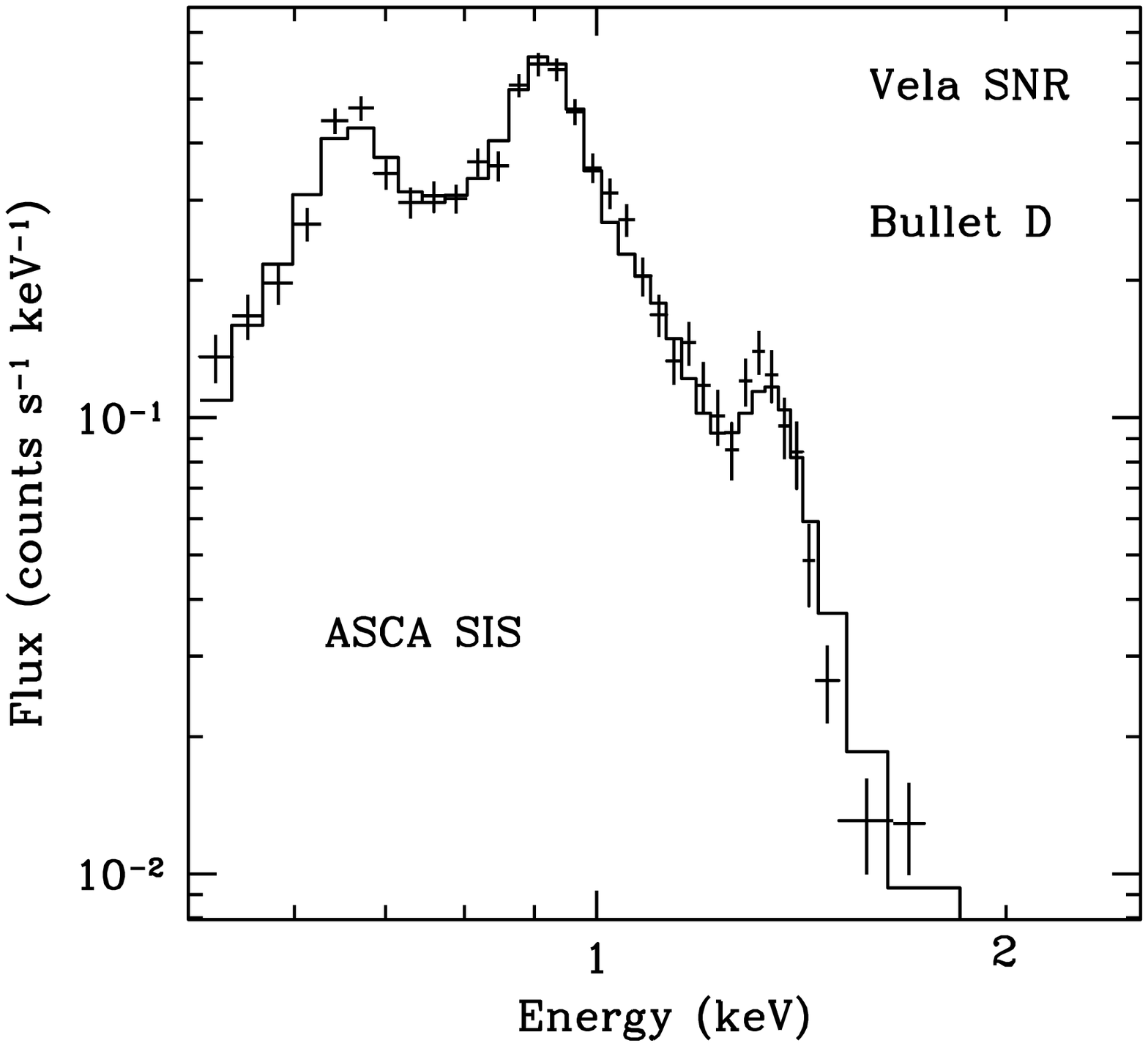,width=2.6in}}
\caption{Left: CO emission ($V_{LSR} = -5 - +20 {\rm\ km\ s}^{-1}$)
in the direction of \snr.  Contours are GIS data. Note the 
Galactic coordinates.
Right: ASCA SIS spectrum from Vela ``Bullet D'' showing thermal nature
of emission, in contrast to nonthermal emission from G266.2$-$1.2.
}
\label{fig2}
\vspace{-0.12in}
\end{figure}

CO data \cite{may} reveal a concentration of
giant molecular clouds -- the Vela Molecular Ridge -- at a distance
of $\sim 1 - 2$~kpc in the direction of Vela.
The column density through the ridge
is in excess of $10^{22}{\rm\ cm}^{-2}$, with the NE rim
of \snr\ (nearest the Galactic Plane) falling along the steeply
increasing column density region
while the western rim lies along a line of much lower
$N_H$ (Fig. 2); the CO column density varies by more
than a factor of 6 between these regions. The lack of a strong variation
in $N_H$ across
\snr\ indicates that the remnant cannot be more distant than the
Ridge. On the other hand, if $N_H$ is
much larger than for the Vela SNR, as the GIS data reported here
indicate, then \snr\ must be as distant as possible consistent with
being in front of most of the Molecular Ridge gas.

In a recent study of optical emission in the vicinity of \snr, 
Redman et al. \cite{redman} discuss a filamentary nebula
lying directly along the edge of the so-called Vela ``bullet D'' region 
\cite{asch2}, just outside the eastern edge of Fig. 1 (in the direction 
of the extended contour), and argue that both structures
represent a breakout region from \snr. However, the
ASCA spectrum of this region (Fig. 2) 
is clearly thermal, with large
overabundances of O, Mg, and Si -- quite in contrast to the nonthermal
spectrum observed for \snr. We conclude that an association between
this region and \snr\ is unlikely.

The central region of the remnant contains a
compact source, which we designate as AX~J0851.9-4617.4,
surrounded by diffuse emission that
extends toward the northwest. The diffuse emission is
well described by a power law of spectral index $\sim 2.0$, again accompanied
by a soft thermal component that we associate with Vela. Given the
nonthermal nature of the shell emission, however, it is quite possible
that the diffuse central emission is associated with emission from the
shell projected along the central line of sight. We note that the spectrum
of this central emission appears harder than that from the rest of
the remnant, perhaps suggesting a plerionic nature, but more sensitive
observations are required to clarify this.

The spectrum of the compact source is rather sparse and can
be adequately described by a variety of models.  A power-law fit
yields a spectral index that is rather steep compared with known pulsars.
The luminosity, on the other hand, is quite reasonable for a young neutron
star if the distance is indeed several kpc, and the observed column density is
compatible with that for the SNR shell. A blackbody model leads to an
inferred surface temperature of $\sim 0.5$~keV with an emitting region
only $\sim 200  d_{\rm kpc} {\rm\ m}$ in radius which could be indicative of
emission from compact polar cap regions of a neutron star, although this
region is somewhat large for such a scenario. It is possible that
the source is associated with a Be star within the position error
circle \cite{asch1}, although the X-ray and optical properties appear
inconsistent for such an interpretation \cite{slane01}.

Recent SAX observations \cite{SAX} have revealed the presence of another
source in the central region of \snr. This source, designated
SAX~J0852.0-4615, has a harder spectrum than AX~J0851.9-4617.4, 
but is considerably 
fainter. If at a distance of $\sim 3$~kpc, its luminosity would be similar
to the Vela pulsar, but this would place the source beyond the Vela
Molecular Ridge. Further observations are required to investigate these
central sources more completely.

\section*{Conclusions}

The ASCA observations of \snr\ reveal that the soft X-ray emission from
this SNR is dominated by nonthermal processes. This brings to three the
number of SNRs in this class, and provides additional evidence for
shock acceleration of cosmic rays in SNRs. Because of the bright and
spatially varying background caused by the Vela SNR, limits on the
thermal emission from \snr\ are difficult to establish. 
However, the ASCA data
suggest a larger column density for this remnant than for Vela,
indicating that \snr\ is at a larger distance,
and perhaps associated with the star formation region located at a
distance of $1 - 2$~kpc. The point source
AX~J0851.9-4617.4 could represent an associated neutron star, 
with the diffuse emission around the source being an associated synchrotron 
nebula, albeit a very faint one. Recent SAX studies
reveal another possible candidate. Higher resolution
X-ray observations of these sources are of considerable importance.

\vspace{0.1in}
\noindent {\bf Acknowledgments.  }
This work was supported in part by NASA through contract NAS8-39073 and
grant NAG5-9106.

\end{document}